\documentclass[12pt]{article}
\usepackage{epsfig}
\usepackage{graphics}
\usepackage{axodraw}
\usepackage{graphicx}
\usepackage{amsmath}
\usepackage{amsfonts}
\usepackage{amssymb}

\newcommand{\comment}[1]{}
\def\be{\begin{equation}}
\def\ee{\end{equation}}
\def\beq{\begin{equation}}
\def\eeq{\end{equation}}
\def\bea{\begin{eqnarray}}
\def\eea{\end{eqnarray}}

\def\gv{g_V}
\def\ga{g_A}
\def\agr{{\mathrm Re~} a_\gamma}
\def\azr{{\mathrm Re~} \Delta a_Z}
\def\agi{{\mathrm Im~} a_\gamma}
\def\azi{{\mathrm Im~} \Delta a_Z}
\def\bgr{{\mathrm Re~} b_\gamma}
\def\bzr{{\mathrm Re~} b_Z}
\def\bgi{{\mathrm Im~} b_\gamma}
\def\bzi{{\mathrm Im~} b_Z}
\def\bgtr{{\mathrm Re~} \tilde b_\gamma}
\def\bztr{{\mathrm Re~} \tilde b_Z}
\def\bgti{{\mathrm Im~} \tilde b_\gamma}
\def\bzti{{\mathrm Im~} \tilde b_Z}

\def\pt{P_T}
\def\ptbar{\overline P_T}
\def\peff{P^{\rm eff}_L}

\def\dsp{\displaystyle }
\def\Re{\mathrm Re~}
\def\Im{\mathrm Im~}

\begin{document}
\begin{flushright}
\end{flushright}

\begin{center}
\boldmath
{\Large \bf Angular distributions as a probe of \\[2mm] 
anomalous $ZZH$ and $\gamma ZH$
interactions \\[4mm] at a linear collider with polarized beams
}
\vskip 1cm
{\large Saurabh D. Rindani and Pankaj Sharma}\\
\smallskip
\smallskip
{\it Theoretical Physics Division, Physical Research Laboratory \\
Navrangpura, Ahmedabad 380 009, India}
\vskip 2cm
{\bf Abstract}
\end{center}

\begin{quote}
We examine the contribution of general $Z^*ZH$ and $\gamma^*ZH$ three-point 
interactions arising from new
physics to the Higgs production 
process $e^+e^- \to HZ$. From Lorentz covariance, each of these 
vertices may be written in terms
of three (complex) form factors,
whose real and imaginary parts together make six independent couplings.  
We take into account possible longitudinal or transverse beam
polarization likely to be available at a linear collider. 
We show how partial cross sections 
and angular asymmetries in suitable combinations with appropriate beam
polarizations, can be used to disentangle various couplings from one
another. A striking result is that using transverse polarization, 
one of the $\gamma ZH$ couplings, 
not otherwise accessible, can be determined
independently of all other couplings.
Transverse polarization also helps in the independent determination of
a combination of two other couplings, in contrast to a combination of four
accessible with unpolarized or longitudinally polarized beams.
We also obtain the sensitivity of the various observables
in constraining the new-physics interactions at a linear collider
operating at a centre-of-mass energy of 500 GeV with longitudinal 
or transverse polarization.

\end{quote}

\vskip 2cm

\section{Introduction}

Despite the dramatic success of the standard model (SM), 
an essential component 
of SM responsible for generating masses in the theory, viz., the Higgs 
mechanism, remains untested. The SM Higgs boson, signalling
symmetry breaking in SM by means of one scalar doublet of $SU(2)$, is
yet to be discovered. A scalar boson with the properties of the SM Higgs
boson is likely to be discovered at the Large Hadron Collider
(LHC). However, there are a number of scenarios beyond the standard
model for spontaneous symmetry
breaking, and ascertaining the mass and other properties of the
scalar boson or bosons is an important task. This task would prove
extremely difficult for LHC. However, scenarios beyond SM, with 
more than just one Higgs doublet, as in the case of the minimal supersymmetric
standard model (MSSM), would be more amenable to discovery at a linear $e^+e^-$
collider operating at a centre-of-mass (cm) energy of 500 GeV. We are at
a stage when the International
Linear Collider (ILC) seems poised to become a reality \cite{LC_SOU}. 

Scenarios going beyond the SM mechanism of symmetry breaking, and
incorporating new mechanisms of CP violation, have also become a
necessity in order to understand baryogenesis which resulted in the
present-day baryon-antibaryon asymmetry in the universe. In a theory
with an extended Higgs sector and new mechanisms of CP violation, the
physical Higgs bosons are not necessarily eigenstates of CP. 
In such a case, the production of a physical Higgs can proceed through
more than one channel, and the interference between two channels can
give rise to a CP-violating signal in the production.

\begin{figure}[htb]
\begin{center}
\epsfig{height=5cm,file=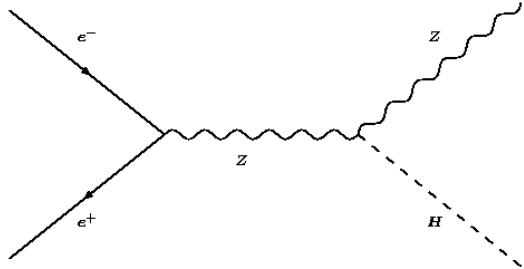}
\end{center}
\caption{Higgs production diagram with an $s$-channel exchange of $Z$
with point-like $ZZH$ coupling.}
\label{fig:vvhptgraph}
\end{figure}

Here we consider in a general model-independent way the production of a
Higgs mass eigenstate $H$ in a possible extension of SM 
through the process $e^+e^- \to HZ$ mediated
by $s$-channel virtual $\gamma$ and $Z$. This is
an important mechanism for the production of the Higgs, the other
important mechanisms being $e^+e^- \to e^+e^- H$ and $e^+e^- \to \nu
\overline \nu H$ proceeding via vector-boson fusion. $e^+e^- \to HZ$ is
generally assumed to get a contribution from a diagram with an 
$s$-channel exchange of $Z$. At the lowest
order, the $ZZH$ vertex in this diagram would be simply a point-like
coupling (Fig. \ref{fig:vvhptgraph}).  Interactions
\begin{figure}[htb]
\begin{center}
\epsfig{height=4cm,file=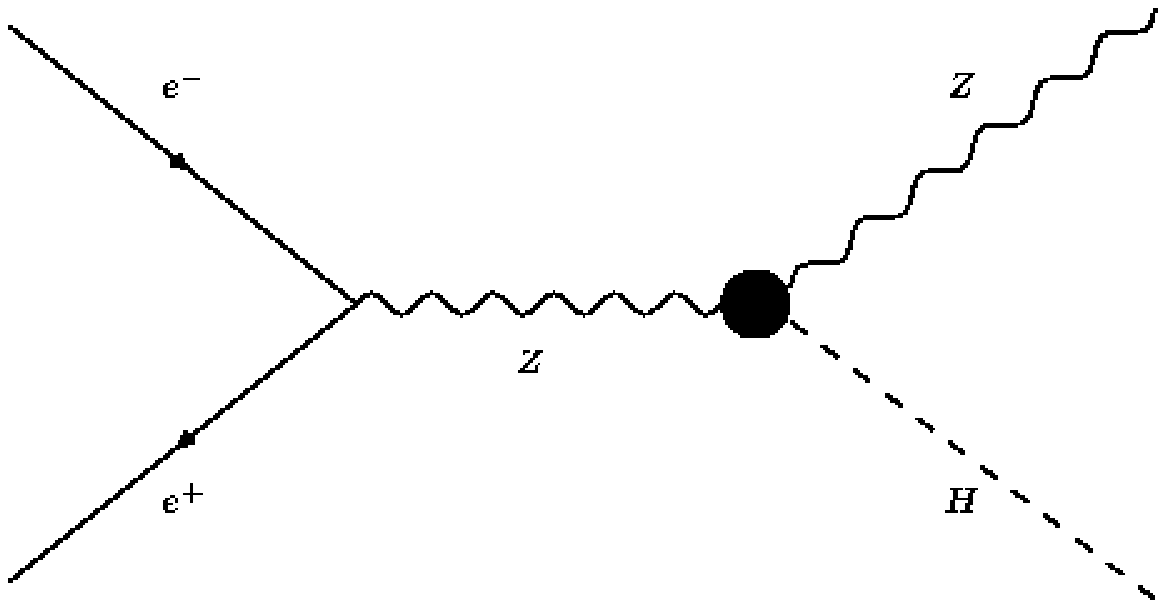}
\end{center}
\caption{Higgs production diagram with an $s$-channel exchange of $Z$
with anomalous $ZZH$ coupling.}
\label{fig:vvhgraph}
\end{figure}
\begin{figure}[htb]
\begin{center}
\epsfig{height=4cm,file=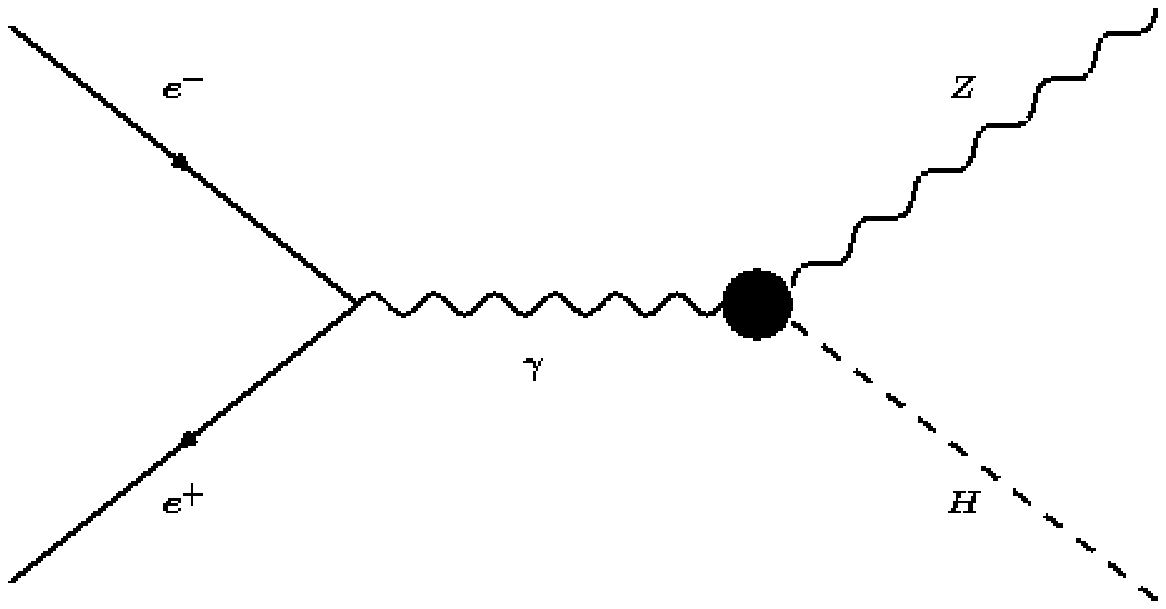}
\end{center}
\caption{Higgs production diagram with an $s$-channel exchange of $Z$
with anomalous $\gamma ZH$ coupling.}
\label{fig:gzhgraph}
\end{figure}
beyond SM can modify this point-like vertex by means of a
momentum-dependent form factor,
as well as by adding more complicated momentum-dependent 
forms of anomalous interactions considered
in \cite{zerwas}-\cite{dutta}. 
The corresponding diagram
is shown in Fig. \ref{fig:vvhgraph}, where the anomalous $ZZH$ vertex is
denoted by a blob. There is also a diagram with a photon
propagator and an anomalous $\gamma ZH$ vertex, which does not occur
in SM at tree level. This is shown in Fig. \ref{fig:gzhgraph} by a blob.
This coupling vanishes in SM at tree level, but can get contributions at
higher order in SM or in extensions of SM. Such  anomalous $\gamma ZH$
couplings were considered earlier in \cite{hagiwara,gounaris,hikk,dutta}.

Refs. \cite{RR1,RR2}  considered a  beyond-SM
contribution represented by a four-point $e^+e^-HZ$ coupling
general enough to include the effects of
the diagrams of Figs. \ref{fig:vvhgraph} and \ref{fig:gzhgraph}, 
as well as additional couplings going beyond $s$-channel exchanges.
By considering appropriate relations between
those form factors and momentum dependencies, we can derive expressions
we consider here. While the four-point coupling is most general, the
dominant contributions are likely to arise from the three-point
couplings considered here.

We write the most general $Z^*ZH$ and $\gamma^*ZH$ couplings consistent
with Lorentz invariance. We do not assume CP
conservation.
We then obtain angular distributions for the $Z$  
arising from the square of amplitude $M_1$ for the diagram in Fig.
\ref{fig:vvhptgraph} with a point-like $ZZH$ coupling, 
together with  the cross terms between $M_1$ and the
amplitude $M_2$ for the diagram in Fig. \ref{fig:vvhgraph}, and the
amplitude $M_3$ for the diagram in Fig. \ref{fig:gzhgraph}. 
We neglect terms quadratic in $M_2$ and $M_3$, 
assuming that the new-physics contribution is small compared to
the dominant SM contribution.
We include the possibility that the beams have polarization, either
longitudinal or transverse.
While we have restricted the actual calculation to SM couplings in
calculating $M_1$, it should be borne in mind that 
in models with more than one Higgs doublet this
amplitude would differ by an overall factor depending on the mixing
among the Higgs doublets. Thus our results are trivially applicable to
such extensions of SM, by an appropriate rescaling of the coupling.

We are thus addressing the question of how well the form factors for the
anomalous $ZZH$ and $\gamma ZH$ 
couplings in $e^+e^-\to HZ$ can be determined from the observation of
$Z$ angular distributions in the presence of unpolarized beams or beams
with either longitudinal or transverse polarizations. This question
taking into account a new-physics contribution which merely modifies the
form of the 
$ZZH$ vertex has been addressed before in several works
\cite{zerwas,skjold,han,biswal,cao}. 
This amounts to assuming that the $\gamma ZH$ couplings are zero or
negligible. Refs. \cite{hagiwara,gounaris,hikk,dutta} do take into account both
$\gamma ZH$ and $ZZH$ couplings. However, they relate both
to coefficients of terms of higher dimensions in an effective
Lagrangian, whereas we treat all couplings as independent of one
another. Moreover, \cite{gounaris} does not discuss effects of beam
polarization. On the other hand, we attempt to seek ways to determine the
couplings completely independent of one another.
Refs. \cite{hikk,dutta} does have a similar approach to ours. They make use of
optimal observables  and consider only longitudinal electron
polarization,
whereas we seek to use simpler observables and
asymmetries constructed out of the $Z$ angular variables, 
and consider the effects of longitudinal and transverse
polarization of both $e^-$ and $e^+$ beams. The authors of \cite{hikk}
also include $\tau$ polarization and $b$-jet charge
identification which we do not require. 

One specific practical
aspect in which our approach differs from
that of the effective Lagrangians is that while the couplings are all
taken to be real in the latter approach, we allow the couplings to be
complex, and in principle, momentum-dependent form factors.

Polarized beams are likely to be available at a linear collider, and
several studies have shown the importance of longitudinal
polarization in reducing backgrounds and improving the sensitivity to
new effects \cite{gudi}. The question of whether transverse beam
polarization, which could be obtained with the use of spin rotators,
would be useful in probing new physics, has been addressed in recent
times in the context of the ILC 
\cite{RR1}-\cite{rizzo}
In earlier work, it has been observed that polarization does not give
any new information about the anomalous $ZZH$ couplings when they are
assumed real \cite{hagiwara}. 
However, the sensitivity can be improved by suitable choice of
polarization. Moreover, polarization can indeed give information
about the imaginary parts of the couplings.
A model-independent approach on kinematic observables in one- and
two-particle final states when longitudinal or transverse beam
polarization is present, which covers our present process, can be found
in \cite{basdrDR}.

In this work, our emphasis has been on simultaneous independent 
determination of couplings, to the extent possible, making use of a
combination of asymmetries and/or polarizations. We have also tried to
consider rather simple observables, conceptually, as well as from an
experimental point of view. With this objective in mind, we use only $Z$
angular distributions without including the polarization or  the 
decay of the $Z$. This
amounts to using the sum of the momenta of the $Z$ decay products. Since
we do not require charge determination, this
has the advantage that one can include both leptonic and hadronic decays
of the $Z$. On the other hand, if a measurement on the Higgs-boson decay
products is made, we can also use the two-neutrino decay
channels of $Z$ since the missing energy-momentum would be fully
determined. 

When all couplings are assumed to be
independent and nonzero, we find that angular asymmetries are linear
combinations of a certain number of anomalous couplings 
(in our approximation of neglecting
terms quadratic in anomalous couplings). By using that many number of
observables, for example, different asymmetries, or the same asymmetry
measured for different beam polarizations, one can solve simultaneous
linear equations to determine the couplings involved.
This is the approach we follow here.
A similar technique of considering combinations of different
polarizations was made use of, for example, in \cite{poulose}.

We find that longitudinal polarization is particularly useful in
achieving our purpose of determining a different combination of
couplings compared to the unpolarized case. As it turns out, the cross
section with transverse
polarization generally provides combinations of the same
couplings as longitudinal polarization. A marked exception is the
angular dependence associated with the coupling $\Im a_\gamma$ (the
couplings are defined in the next section) -- it is
possible to use an azimuthal asymmetry which depends entirely on this
coupling when the beams are transversely polarized, and its measurement
would determine this coupling directly. Unpolarized or longitudinally
polarized beams provide no access to $\Im a_\gamma$.
Another azimuthal asymmetry in the presence of transverse polarization 
helps to isolate a combination of two
couplings $\agr$ and $\azr$ out of the four which contribute to the
differential cross section with longitudinal polarization.
 
In the next section we write down the possible model-independent
$ZZH$ and $\gamma ZH$ couplings. In Section 3, we obtain the angular
distributions arising from these couplings in the presence of beam
polarization. Section 4 deals with asymmetries which can be used
for separating various form factors and Section 5 describes the numerical
results.  Section 6
contains our conclusions and a discussion.

\section{\boldmath Form factors for the process $e^+e^- \to HZ$}

Assuming Lorentz invariance, the
general structure for the vertex corresponding to the process
$V_\mu^* (k_1) \to Z_\nu (k_2) H$, where $V\equiv \gamma$ or $Z$, can
be written as \cite{skjold,han,biswal,hikk}
\be\label{couplings}
\Gamma_{\mu\nu} = g_Vm_Z \left[ a_V\,g_{\mu\nu} + \frac{b_V}{m_Z^2}\,
	(k_{1\nu}k_{2\mu} - g_{\mu\nu}k_1\cdot k_2) + 
	\frac {\tilde b_V}{m_Z^2}\,\epsilon_{\mu\nu\alpha\beta}
	k_1^\alpha k_2^\beta\right],
\ee
where $a_V$, $b_V$ and $\tilde b_V$, are form factors, which are in
general complex. We have omitted terms proportional to $k_{1\mu}$ and
$k_{2\nu}$, which do not contribute to the process $e^+e^- \to HZ$ in
the limit of vanishing electron mass. The constant
$g_Z$ is chosen to be $g/\cos\theta_W$, 
so that $a_Z=1$ for SM. $g_\gamma$ is chosen to be $e$. 
Of the interactions in (\ref{couplings}), the terms with 
$\tilde b_Z$ and $\tilde b_\gamma$ are CP 
violating, whereas the others are CP conserving.
Henceforth we will write $a_Z=1+ \Delta a_Z$, $\Delta a_Z$ being the
deviation of $a_Z$ from its tree-level SM value. 
The other form factors are vanishing in SM at tree level. Thus the above
``couplings,'' which are deviations from the tree-level SM values, could
arise from loops in SM or from new physics beyond SM. We could of course
work with a set of modified couplings where the anomalous couplings denote
deviations from the tree-level values in a specific extension of the SM
model, like a concrete two-Higgs doublet  model. The corresponding
modifications are trivial to incorporate.

\comment{
It may be appropriate to contrast our approach with the usual effective
Lagrangian approach. In the latter approach, it is
assumed that SM is an effective theory which is valid up to a cut-off scale
$\Lambda$. The new physics occurring above the scale of the cut-off may
be parametrized by higher-dimensional operators, appearing with powers
of $\Lambda$ in the denominator. These when added to the SM Lagrangian
give an effective low-energy Lagrangian
where, depending on the scale of the momenta involved, one
includes a range of higher-dimensional operators up to a certain maximum
dimension. Our effective theory is not a low-energy limit, so that the
form factors we use are functions of momentum not restricted to low
powers. Thus, we have introduced $m_Z$  rather than a cut-off scale, 
just to make the form factors dimensionless. 
}

The expression for the amplitude for the process 
\begin{equation}\label{process}
e^- (p_1) + e^+ (p_2) \to Z^\alpha (q) + H(k),
\end{equation}
arising from the SM 
diagram of Fig. \ref{fig:vvhptgraph} with a point-like $ZZH$ vertex, is 
\begin{equation}\label{smamp}
M_{\rm SM} =- \frac{e^2}{4\sin^2\theta_W \cos^2\theta_W}
\frac{m_Z}{s-m_Z^2} \overline v (p_2) \gamma^\alpha (\gv - \gamma_5 \ga)
u(p_1),
\end{equation}
where the vector and axial-vector couplings of the $Z$ to electrons are
given by
\beq\label{gvga}
g_V^e= -1 + 4 \sin^2\theta_W,\; g_A^e=-1,
\eeq
and $\theta_W$ is the weak mixing angle. 

\section{Differential cross sections}

We now obtain the differential cross section for the process
(\ref{process}) keeping the pure SM contribution, and the interference
between the SM amplitude of Fig. \ref{fig:vvhptgraph} 
and the amplitudes with anomalous $\gamma ZH$ and $ZZH$
couplings  of Figs.  \ref{fig:vvhgraph} and  \ref{fig:gzhgraph},
respectively.
We ignore terms
bilinear in the anomalous couplings, assuming that the new-physics
contribution is small. We treat the two cases of longitudinal and
transverse polarizations for the electron and positron beams separately.
We neglect the mass of the electron. 

Helicity amplitudes for the process were obtained earlier in the context
of an effective Lagrangian approach \cite{gounaris,hagiwara,hikk,dutta},
and could be made use of for obtaining the differential cross section
for the case of longitudinal polarization, and, with less ease, for the
case of transverse polarization.
We have used instead trace techniques employing the
symbolic manipulation program `FORM' \cite{form}.

Note that though we have used SM couplings for the leading contribution,
it is trivial to modify these by overall factors for cases of other
models (like two-Higgs-doublet models). Our expressions are not, however, 
applicable for the case when the Higgs is a pure pseudoscalar in models
conserving CP, since in that case, the SM-like lowest-order couplings
are absent. 

We choose the $z$ axis to be the direction of the $e^-$ momentum, and
the $xz$ plane to coincide with the $HZ$ production plane in the case
when the initial beams are unpolarized or longitudinally polarized. 
The positive $x$
axis is chosen, in the case of transverse polarization, to be along the
direction of the $e^-$ polarization. 
We then define $\theta$ and $\phi$
to be the polar and azimuthal angles of the momentum $\vec q$ of the $Z$.
We use the convention $\epsilon^{0123}=+1$ for the Levi-Civita tensor.

\subsection{Angular distributions for longitudinal polarization}

The angular distribution for the process (\ref{process}) with 
longitudinal polarizations $P_L$ and $\overline P_L$ respectively 
of the $e^-$ and
$e^+$  beams may be written as
\begin{equation}\label{longcs}
  \frac{d\sigma_L}{d\Omega}=\left(1-P_L\overline{P}_L \right) 
[A_L +B_L\sin^2\theta + C_L\cos\theta ],
\end{equation}
where $A_L$, $B_L$, $C_L$ are further written in terms of contributions
from SM alone (superscript ``SM''), interference between SM and $ZZH$
terms (superscript $Z$), and interference between SM and $\gamma ZH$
(superscript $\gamma$):
\be\label{coeffs}
A_L = A_L^{\rm SM} + A_L^{Z} + A_L^{\gamma},
\ee 
\be
B_L = B_L^{\rm SM} + B_L^{Z} + B_L^{\gamma},
\ee 
\be
C_L = C_L^{Z} + C_L^{\gamma}.
\ee 
In case of $C_L$, which is the coefficient of a CP-odd term, 
there is no contribution from SM.
The expressions for the various terms used above are as follows.
\comment{
A_L^{\rm SM} = 
		(g_V^{e2} + g_A^{e2}-2g_V^e g_A^e P_L^{\rm
		eff})K_L^{\mathrm SM},\;
}
\be
A_L^{\mathrm SM} = 
		B_L^{\rm SM} \frac {2m_Z^2} {\vert \vec q \vert^2} = 
		(g_V^{e2} + g_A^{e2}-2g_V^e g_A^e P_L^{\rm
		eff})K^{\mathrm SM},\;
C_L^{\rm SM} = 0,
\ee
\comment{
	
		(g_V^{e2} + g_A^{e2}-2g_V^e g_A^e P_L^{\rm
		eff})K_L^{\mathrm SM},
\be
B_L^{\rm SM} = 
		\frac{\alpha^2\vert \vec q \vert }{2 \sqrt{s}\sin^4 2\theta_W}
		\frac{\vert
		\vec q \vert^2} {2(s-m_Z^2)^2}
		(g_V^{e2} + g_A^{e2}-2g_V^e g_A^e P_L^{\rm eff}),
\ee
}
where
\be
	K^{\mathrm SM} =
		\frac{\alpha^2\vert \vec q \vert }{2 \sqrt{s}\sin^4 2\theta_W}
		\frac{m_Z^2} {(s-m_Z^2)^2}.
\ee
We also have
\be
A_L^{Z}=
		\displaystyle 2 \left(\Re\Delta a_Z + \Re b_Z
                \frac{\sqrt{s}q^0}{m_Z^2}\right) 
		(g_V^{e2} + g_A^{e2}-2g_V^e g_A^e P_L^{\rm eff})
		K^{\mathrm SM},
\ee
\be
B_L^{Z} = 
		\displaystyle 2 \Re\Delta a_Z 
                \frac{\vert \vec q \vert^2}{2 m_Z^2}  
		(g_V^{e2} + g_A^{e2}-2g_V^e g_A^e P_L^{\rm eff})
		K^{\mathrm SM},
\ee
\be
C_L^{Z} = 
               2 \Im\tilde b_Z\frac{\sqrt{s}\vert\vec q \vert }{m_Z^2}
                \left((g_V^{e2} + g_A^{e2}) P_L^{\rm eff}
			-2g_V^e g_A^e\right)
		K^{\mathrm SM}.
\ee
\comment{
\be
\begin{array}{lcc}
A_L^{Z}&=&\displaystyle
		\frac{\alpha^2\vert \vec q \vert }{\sqrt{s}\sin^4 2\theta_W}
		\frac{m_Z^2}{(s-m_Z^2)^2}
		(g_V^{e2} + g_A^{e2}-2g_V^e g_A^e P_L^{\rm eff})
		\\&& \displaystyle
		\times \left(\Re\Delta a_Z + \Re b_Z
		\frac{\sqrt{s}q^0}{m_Z^2}\right),
\end{array}
\ee
\be
B_L^{Z} = 
		\frac{\alpha^2\vert \vec q \vert }{ \sqrt{s}\sin^4 2\theta_W}
		\frac{\vert
		\vec q \vert^2} {2(s-m_Z^2)^2}\Re\Delta a_Z
		(g_V^{e2} + g_A^{e2}-2g_V^e g_A^e P_L^{\rm eff}),
\ee
\be
C_L^{Z} = 
		\frac{\alpha^2\vert \vec q \vert }{ \sqrt{s}\sin^4 2\theta_W}
                \frac{m_Z^2}{(s-m_Z^2)^2}
                \Im\tilde b_Z\frac{\sqrt{s}\vert\vec q \vert }{m_Z^2}
                \left((g_V^{e2} + g_A^{e2}) P_L^{\rm eff}
			-2g_V^e g_A^e\right),
\ee
} 
Next,
\be
A_L^{\gamma} = \displaystyle 
		\left(\Re a_{\gamma} + 
		\Re b_{\gamma}\frac{\sqrt{s}q^0}{m_Z^2}\right)
		(g_V^e - g_A^e P_L^{\rm eff}) K^\gamma ,
\ee
\be
B_L^{\gamma} = 
		\Re a_{\gamma}
                \frac{\vert \vec q \vert^2}{2 m_Z^2}  
		(g_V^e - g_A^e P_L^{\rm eff})K^\gamma ,\;
C_L^{\gamma} = 
		\frac{\sqrt{s}\vert\vec q \vert }{m_Z^2}
                \Im\tilde b_{\gamma}
                \left( g_A^{e} - g_V^e P_L^{\rm eff} \right) K^\gamma ,
\ee
where 
\be 
	K^\gamma = \dsp
		\frac{\alpha^2\vert \vec q \vert }{ \sqrt{s}\sin^2 2\theta_W}
		\frac{m_Z^2}{s(s-m_Z^2)}.
\ee
\comment{
\be
\begin{array}{lcl}
A_L^{\gamma}& =&\displaystyle 
		\frac{\alpha^2\vert \vec q \vert }{ \sqrt{s}\sin^2 2\theta_W}
		\frac{m_Z^2}{s(s-m_Z^2)}
		(g_V^e - g_A^e P_L^{\rm eff})
		\\&&\displaystyle\times
		\left(\Re a_{\gamma} + 
		\Re b_{\gamma}\frac{\sqrt{s}q^0}{m_Z^2}\right),
\end{array}
\ee
\be
B_L^{\gamma} = 
		\frac{\alpha^2\vert \vec q \vert }{ \sqrt{s}\sin^2 2\theta_W}
		\frac{\vert\vec q\vert^2}
		{2s(s-m_Z^2)}
		\Re a_{\gamma}
		(g_V^e - g_A^e P_L^{\rm eff}),
\ee
\be
C_L^{\gamma} = 
		\frac{\alpha^2\vert \vec q \vert }{ \sqrt{s}\sin^2 2\theta_W}
		\frac{m_Z^2}{s(s-m_Z^2)}
		\frac{\sqrt{s}\vert\vec q \vert }{m_Z^2}
                \Im\tilde b_{\gamma}
                \left( g_A^{e} - g_V^e P_L^{\rm eff} \right),
\ee
} 

In the above, we have used the effective polarization
\be\label{pleff}
P_L^{\rm eff} = \frac{P_L - \overline P_L}{1 - P_L \overline P_{L}}.
\ee
The expressions for the $Z$ energy $q^0$ and the magnitude of its 
three-momentum $\vert \vec q \vert$ are
\be
q^0 = \frac{s+m_Z^2-m_H^2}{2\sqrt{s}}, \;
\vert \vec q \vert = \frac{\sqrt{s^2 + (m_Z^2 - m_H^2)^2 - 2 s(m_Z^2+m_H^2)}}
			{2\sqrt{s}}.
\ee 

Immediate inferences from these expressions are: (i) If the six
coefficients $A_L^{\gamma,Z}$, $B_L^{\gamma,Z}$ and $C_L^{\gamma,Z}$
could be determined independently using angular distributions and
polarization, it would be possible to determine the six anomalous couplings
$\agr, \azr, \bgr, \bzr, \bgti$ and $\bzti$. 
(ii)
Imaginary parts of $a_{\gamma}$, $\Delta a_Z$, $b_{\gamma}$, $b_Z$, 
and real parts
of $\tilde b_{\gamma}$, $\tilde b_Z$ 
do not contribute to the angular distributions
at this order, and hence remain undetermined.  
(iii) Numerically
$g_V^e$ is small (about $-0.12$ for $\sin^2\theta_W = 0.22$), while
$g_A^e=-1$. Hence, in the absence of polarization, from among the anomalous
contributions, the terms $A_L^Z$, $B_L^Z$ and $C_L^{\gamma}$ dominate
over the others. If these coefficients are determined from angular
distributions, it would be possible to measure $\azr, \bzr$
and $\Im \tilde b_\gamma$ with greater sensitivity.
On the other hand, there
would be very low sensitivity to the
remaining couplings, viz.,  $\agr,\bgr$
and $\Im \tilde b_Z$. (iv) Within the combinations of couplings which
appear in $A_L^\gamma$ and $A_L^Z$, the contributions of $\bgr$ and
$\bzr$
are enhanced because of the factor $\sqrt{s}q^0/m_Z^2$ multiplying them.
This improves their sensitivity. 
(v) With longitudinal polarization turned on, with a
reasonably large value of  $\peff$, the coefficients $C_L^Z$, $A_L^\gamma$
and $B_L^\gamma$ would become significant. In that case, the sensitivity
to $\Re a_{\gamma}$, $\Re b_{\gamma}$
and $\Im \tilde b_Z$ would be improved. (vi) In view of (v), it is
clear that a combination of angular distributions for the polarized and
unpolarized cases will help in disentangling the different couplings.

\subsection{Angular distributions for transverse polarization}

For the transverse case, we take the $e^-$ polarization to be along the
$x$ axis 
and that of the $e^+$ in the $xy$ plane, making an angle of $\delta$ 
with the
$x$ axis, so that $\delta=0$ corresponds to parallel  $e^-$ and
$e^+$ transverse polarizations.
The expression for the cross section with transverse polarization $P_T$ for
the $e^-$ beam and $\overline P_T$ for the $e^+$ beam is
\begin{equation}\label{sigtran}
\begin{array}{lll}
\dsp
  \frac{d\sigma_T}{d\Omega}&=&
\left[A_T +B_T\sin^2\theta + C_T\cos\theta \right.\\
& &+\left. P_T\overline P_T \sin^2\theta
\left\{D_T \cos(2\phi - \delta) + E_T \sin (2\phi - \delta) \right\}
\right],
\end{array}
\end{equation}
where $A_T$, $B_T$, $C_T$, $D_T$ and $E_T$
 are further written in terms of contributions
from SM alone (superscript ``SM''), interference between SM and $ZZH$
terms (superscript $Z$), and interference between SM and $\gamma ZH$
(superscript $\gamma$), in exact analogy with expressions for $A_L$,
$B_L$ and $C_L$ given earlier for the longitudinal polarization case. 
The expressions for the separate contributions for these coefficients
are as follows.
\comment{
\be
A_T^{\rm SM} = 
		\frac{\alpha^2\vert \vec q \vert }{2 \sqrt{s}\sin^4 2\theta_W}
		\frac{m_Z^2}{(s-m_Z^2)^2}
		(g_V^{e2} + g_A^{e2}),
\ee
\be
B_T^{\rm SM} = 
		\frac{\alpha^2\vert \vec q \vert }{2 \sqrt{s}\sin^4 2\theta_W}
		\frac{\vert
		\vec q \vert^2} {2(s-m_Z^2)^2}
		(g_V^{e2} + g_A^{e2}),
\ee
} 
\be
		A_T^{\rm SM} = 
			B_T^{\rm SM}  \frac {2m_Z^2} {\vert \vec q \vert^2}
	=	(g_V^{e2} + g_A^{e2})K^{\mathrm SM},
\ee
\be
C_T^{\rm SM} = 0,\; 
D_T^{\rm SM} = 
		\frac{\vert \vec q \vert^2}{2m_Z^2}
		(g_V^{e2} - g_A^{e2})K^{\mathrm SM}, \;
E_T^{\rm SM} = 0,
\ee
\comment{
\be
\begin{array}{lcc}
A_T^{Z}&=&\displaystyle
		\frac{\alpha^2\vert \vec q \vert }{ \sqrt{s}\sin^4 2\theta_W}
		\frac{m_Z^2}{(s-m_Z^2)^2}
		(g_V^{e2} + g_A^{e2})
		\\&& \displaystyle
		\times \left(\Re\Delta a_Z +
		\Re b_Z\frac{\sqrt{s}q^0}{m_Z^2}\right),
\end{array}
\ee
} 
\be
A_T^{Z} =
		2(g_V^{e2} + g_A^{e2})
		\left(\Re\Delta a_Z +
		\Re b_Z\frac{\sqrt{s}q^0}{m_Z^2}\right)K^{\mathrm
SM},\;
\ee
\be
B_T^{Z} = 
		2\frac{\vert \vec q \vert^2} {2m_Z^2}\Re\Delta a_Z
		(g_V^{e2} + g_A^{e2})K^{\mathrm SM},\;
C_T^{Z} = 
                2{\mathrm Im}\tilde b_Z\frac{\sqrt{s}\vert\vec q \vert }{m_Z^2}
			2g_V^e g_A^eK^{\mathrm SM},
\ee
\comment{
\be
B_T^{Z} = 
		\frac{\alpha^2\vert \vec q \vert }{ \sqrt{s}\sin^4 2\theta_W}
		\frac{\vert
		\vec q \vert^2} {2(s-m_Z^2)^2}\Re\Delta a_Z
		(g_V^{e2} + g_A^{e2}),\;
C_T^{Z} = 
                {\mathrm Im}\tilde b_Z\frac{\sqrt{s}\vert\vec q \vert }{m_Z^2}
			2g_V^e g_A^eK^{\mathrm SM},
\ee
\be
C_T^{Z} = 
		\frac{\alpha^2\vert \vec q \vert }{ \sqrt{s}\sin^4 2\theta_W}
                \frac{m_Z^2}{(s-m_Z^2)^2}
                \Im\tilde b_Z\frac{\sqrt{s}\vert\vec q \vert }{m_Z^2}
                \left(
			2g_V^e g_A^e\right),
\ee
\be
D_T^{Z} = 
		\frac{\alpha^2\vert \vec q \vert }{ \sqrt{s}\sin^4 2\theta_W}
		\frac{\vert
		\vec q \vert^2} {2(s-m_Z^2)^2}
		(g_V^{e2} - g_A^{e2})(-\Re\Delta a_Z),
\ee
 } 
\be
D_T^{Z} = 
		2\frac{\vert
		\vec q \vert^2} {2m_Z^2}
	(-{\mathrm Re}\Delta a_Z)
		(g_V^{e2} - g_A^{e2})
	K^{\mathrm SM}, \;
E_T^Z = 0,
\ee
\comment{
\be
\begin{array}{lcl}
A_T^{\gamma}& =&\displaystyle 
		\frac{\alpha^2\vert \vec q \vert }{ \sqrt{s}\sin^2 2\theta_W}
		\frac{m_Z^2}{s(s-m_Z^2)}
		(g_V^e )
		\\&&\displaystyle\times
		\left({\mathrm Re} a_{\gamma} + 
		{\mathrm Re}b_{\gamma}\frac{\sqrt{s}q^0}{m_Z^2}\right),
\end{array}
\ee
\be
B_T^{\gamma} = 
		\frac{\alpha^2\vert \vec q \vert }{ \sqrt{s}\sin^2 2\theta_W}
		\frac{\vert
		\vec q \vert^2} {2s(s-m_Z^2)}
		\Re a_{\gamma}
		(g_V^e  ),
\ee
} 
\be
A_T^{\gamma} = 
		\left({\mathrm Re} a_{\gamma} + 
		{\mathrm Re}b_{\gamma}\frac{\sqrt{s}q^0}{m_Z^2}\right)(g_V^e ) 
		K^\gamma,\;
\ee
\be
B_T^{\gamma} = 
		\frac{\vert \vec q \vert^2} {2m_Z^2}
		{\mathrm Re} a_{\gamma} (g_V^e  )  K^\gamma,\;
C_T^{\gamma} = 
		\frac{\sqrt{s}\vert\vec q \vert }{m_Z^2}
                {\mathrm Im}\tilde b_{\gamma}
                \left( g_A^{e} \right)  K^\gamma,
\ee
\be
D_T^{\gamma} = \frac{\vert \vec q \vert^2} {2m_Z^2}  
		{\mathrm Re} a_{\gamma} (-g_V^e  )  K^\gamma,\;
E_T^{\gamma} = \frac{\vert \vec q \vert^2} {2m_Z^2}  
		{\mathrm Im} a_{\gamma} (g_A^e  )K^\gamma.
\ee
\comment{
\be
C_T^{\gamma} = 
		\frac{\alpha^2\vert \vec q \vert }{ \sqrt{s}\sin^2 2\theta_W}
		\frac{m_Z^2}{s(s-m_Z^2)}
		\frac{\sqrt{s}\vert\vec q \vert }{m_Z^2}
                \Im\tilde b_{\gamma}
                \left( g_A^{e} \right),
\ee
\be
D_T^{\gamma} = 
		\frac{\alpha^2\vert \vec q \vert }{ \sqrt{s}\sin^2 2\theta_W}
		\frac{\vert
		\vec q \vert^2} {2s(s-m_Z^2)}
		\Re a_{\gamma}
		(-g_V^e  ),
\ee
\be
E_T^{\gamma} = 
		\frac{\alpha^2\vert \vec q \vert }{ \sqrt{s}\sin^2 2\theta_W}
		\frac{\vert
		\vec q \vert^2} {2s(s-m_Z^2)}
		\Im a_{\gamma}
		(g_A^e  ).
\ee
} 
Taking a look at the above equations, one can infer the following: (i)
For studying any effects dependent on transverse polarization, and
therefore, of the azimuthal distribution of the $Z$, both
electron and positron beams have to be polarized. (ii) If the azimuthal
angle $\phi$ of $Z$ is integrated over, there is no difference between
the transversely polarized and unpolarized cross sections \cite{hikasa}. 
Thus the
usefulness of transverse polarization comes from the study of nontrivial
$\phi$ dependence. 
(iii) A glaring
advantage of using transverse polarization would be to determine
$\Im a_\gamma$ from the $\sin(2\phi - \delta)$ dependence of the
angular distribution. It can be seen that $E_T$ receives contribution
only from $E_T^\gamma$, which determines $\Im a_\gamma$ independently of
any other coupling. Moreover, $\Im a_\gamma$ does not contribute to
unpolarized or longitudinally polarized cases. 
(iv) The $\cos(2\phi - \delta)$ dependence of the
angular distribution (the $D_T$ term) determines a 
combination only of the couplings $\Re \Delta a_Z$ and $\Re a_\gamma$. 
On the other hand, in the case of unpolarized or longitudinally
polarized beams the coefficient  
$B_L$ does depend only on $\agr$ and $\azr$, and if measured, can give
information on $\agr$ and $\azr$ independently of $\bgr$ and
$\bzr$. However, there is no simple
asymmetry which allows $B_L$ to be measured separately from $A_L$, which
depends on a combination of all four of $\agr$, $\azr$, $\bgr$ and
$\bzr$.
Of course, it is in principle possible to separate $A_L$ and $B_L$ 
using either a fit or using cross sections integrated over different ranges. 
The latter approach has been used in Sec. 5.1 to obtain simultaneous
limits on all of $\agr$, $\azr$, $\bgr$ and
$\bzr$.
(v) The real parts of the
CP-violating couplings $\tilde b_Z$ and $\tilde b_\gamma$ remain
undetermined with either longitudinal or transverse polarization. (vi)
$\Im \Delta a_Z$ also remains undetermined.

We now examine how the angular distributions in the presence of
polarizations may be used to determine the various form factors.

\section{Polarization and Angular asymmetries} 

In this section we discuss observables like partial cross sections and
angular asymmetries which can be used to determine the anomalous
couplings.

One of the simplest observables is a partial cross section, i.e., the
differential cross section integrated over all azimuthal angles, but
over a limited range in $\theta$. A cut-off in the forward and backward
directions is natural for avoiding the beam pipe, and so this would be
an obvious cut on $\theta$. Such a partial cross section would get
contribution from the SM terms as well as a
linear combination of the real parts of anomalous couplings $\Re \Delta
a_Z$, $\Re a_\gamma$, $\Re b_Z$ and $\Re b_\gamma$, provided the range in
$\theta$ is forward-backward symmetric. The result for unpolarized and
transversely polarized beams would be identical \cite{hikasa}.

On the other hand, with longitudinal polarization, the partial cross
section depends on a different linear combination of the real parts of
the anomalous couplings. Thus, combining results of the measurement
with unpolarized beams with those of the measurement with longitudinally
polarized beams with $e^-$ and $e^+$ polarizations of the same sign or
opposite signs would give three relations with which to constrain the
four couplings.

The expression for the partial cross sections in the longitudinal
polarization case, in terms of the coefficients
$A_L$ and $B_L$ used in the differential cross section is
\be\label{partialcs}
\sigma_L ( \theta_0) = (1 - P_L\overline P_L)4 \pi \cos\theta_0 \left[
			A_L + \left(1 - \frac{1}{3} \cos^2\theta_0\right)
			B_L\right],
\ee 
where $\theta_0$ is the cut-off angle.
As mentioned earlier, the partial cross section in the case of
transverse polarization is the same as that for unpolarized beams.

The terms proportional to 
$\cos\theta$ can be determined using a simple forward-backward asymmetry:
\beq\label{fbasy}
A_{\rm FB}(\theta_0) =  \frac{1}{\sigma(\theta_0)} 
\left[
\int^{\pi/2}_{\theta_0} \frac{d\sigma}{d\theta}d\theta
-\int^{\pi-\theta_0}_{\pi/2} \frac{d\sigma}{d\theta}d\theta
\right],
\eeq
where 
\beq
\sigma(\theta_0) = \int^{\pi-\theta_0}_{\theta_0}
\frac{d\sigma}{d\theta}d\theta,
\eeq
and $\theta_0$ is a cut-off in the forward and backward directions
which could be
chosen to optimize the sensitivity.

The expression for $A^L_{\rm FB}(\theta_0)$ for longitudinal
polarization is
\be\label{fbasyexp}
A^L_{\rm FB}(\theta_0) = \frac{C_L \cos\theta_0}
				{2\left[ A_L^{\mathrm SM} + B_L^{\mathrm
SM}\left(1 - \frac{1}{3}
				\cos^2\theta_0\right)\right]},
\ee
where we have used only the SM cross section in the denominator because
we work to linear order in the anomalous couplings.
This asymmetry is odd under CP and is 
proportional to $C$ and therefore to a combination of $\Im \tilde b_Z$
and $\Im \tilde b_\gamma$. This combination is dependent on
the degree of  (longitudinal) polarization, and is therefore sensitive
to polarization. The asymmetry for transverse
polarization is the same as that for zero polarization.
 
It should be noted that only
imaginary parts of couplings  enter. This is related to the
fact that the CP-violating asymmetry $A_{\rm FB}(\theta_0)$ is odd
under naive CPT. It follows that for it to have a non-zero value, the
amplitude should have an absorptive part \cite{sdrCP}.  

We now treat the cases of longitudinally and transversely polarized
beams separately.

\vskip .2cm
\noindent {\it Case (a) Longitudinal polarization:}

The forward-backward asymmetry of eq. (\ref{fbasy}) in the presence of
longitudinal polarization,  which we denote by $A_{\rm
FB}^{\rm L}(\theta_0)$, determines a different
combination of the same couplings $\Im \tilde b_Z$ and $\Im \tilde b_\gamma$.
Thus observing asymmetries with and
without polarization, the two imaginary parts can be determined
independently.

In the same way, a combination of the cross section for the unpolarized
and longitudinally polarized beams can be used to determine two
different combinations of the
remaining couplings which appear in (\ref{longcs}). However, one can get
information only on the real parts of $\Delta a_Z,b_Z,a_\gamma,b_\gamma$, not
their imaginary parts.

\vskip .2cm
\noindent{\it Case (b) Transverse polarization:}
\nopagebreak

In the case of the angular distribution with transversely polarized
beams, there is a dependence on the azimuthal angle $\phi$ of the $Z$.
Thus, in addition to $\phi$-independent terms which are the same as
those in the unpolarized case, there are terms with factors $\sin^2\theta
\cos 2 \phi$ and  $\sin^2\theta \sin  2 \phi$.
The $\phi$-dependent terms 
occur with the factor of $\pt\ptbar$.
Thus, both beams need
to have transverse polarization for a nontrivial azimuthal dependence. 
\comment{
We find that that with
the possibility of flipping transverse polarization of one beam, 
it is possible to
examine 4 types of angular asymmetries if the SM contribution is
nonzero, and 2 independent types if the SM contribution is vanishing.
Each angular asymmetry would enable the determination of a different 
combination of couplings. 
}

We can define an azimuthal asymmetry which can be 
used to separate out $\Im(a_{\gamma})$:
\beq\label{atfb}
\begin{array}{lcrrl}
A^{\rm T}(\theta_0)& =& \dsp\frac{1}{\sigma^{\mathrm SM}_T(\theta_0)}
		\!\!&\dsp\left[
\int^{\pi-\theta_0}_{\theta_0}d\theta \right.
&\left.
\dsp\left(\int^{\pi/2}_{0} d\phi  - \int^{\pi}_{\pi/2} d\phi 
\right. \right.  \\
&&&& \left.\left.\dsp
+ \int^{3\pi/2}_{\pi} d\phi  - \int^{2\pi}_{3\pi/2} d\phi \right)
\dsp\frac{d\sigma_T}{d\theta d\phi}
\right],
\end{array}
\eeq
Next, we define an asymmetry which separates out a linear combination
of $\Re \Delta a_Z$ and $\Re a_{\gamma}$ as follows:
\beq\label{atfb2}
\begin{array}{lcrrl}
A^{'\rm T}(\theta_0)& =& \dsp\frac{1}{\sigma^{\mathrm SM}_T(\theta_0)}
		\!\!&\dsp\left[
\int^{\pi-\theta_0}_{\theta_0}d\theta \right.
&\left.
\dsp\left(\int^{\pi/4}_{-\pi/4} d\phi  - \int^{3\pi/4}_{\pi/4} d\phi 
\right. \right.  \\
&&&& \left.\left.\dsp
+ \int^{5\pi/4}_{3\pi/4} d\phi  - \int^{7\pi/4}_{5\pi/4} d\phi \right)
\dsp\frac{d\sigma_T}{d\theta d\phi}
\right],
\end{array}
\eeq
The integrals in the above may be evaluated to yield
\beq\label{atfbexp}
A^{\rm T}(\theta_0)= 
			\frac{2}{\pi}  \pt\ptbar 
		\frac{(D_T \sin\delta + E_T \cos\delta ) 
			\left(1-\frac{1}{3} \cos^2 \theta_0 \right) }
			{A_T^{\rm SM} + B_T^{\rm SM} 
		\left(1-\frac{1}{3} \cos^2 \theta_0 \right)},
\eeq
and
\beq\label{atfb2exp}
A^{' \rm T}(\theta_0) =   
			\frac{2}{\pi}  \pt\ptbar 
			\frac{(D_T \cos\delta - E_T \sin\delta ) 
			\left(1-\frac{1}{3} \cos^2 \theta_0 \right) }
			{A_T^{\rm SM} + B_T^{\rm SM} 
		\left(1-\frac{1}{3} \cos^2 \theta_0 \right)},
\eeq
where we use only the SM cross section in the denominators, since we
work to first order in anomalous couplings.
The simplest scenario is when $\delta = 0$ or $\pi$. In that case,
we see that the two asymmetries $A^{\rm T}$ and 
$A^{' \rm T}$ can measure, respectively, $\Im a_\gamma$ and a combination 
of $\Re \Delta
a_Z$ and $\Re a_\gamma$.
The former is odd under naive time reversal, whereas the latter is even.
The CPT theorem then implies that these would be respectively dependent
on real and imaginary parts of form factors \cite{sdrCP}. 
We thus have the important result that a measurement of
$A^{\rm T}(\theta_0)$ when the electron and positron polarizations are
parallel to each other directly gives us a measurement of 
$\Im a_\gamma$, which cannot be measured without the use of 
transverse polarization. This, in the present context, is the most
important use of transverse polarization\footnote{It may be mentioned
that in \cite{RR1}, where $e^+e^-HZ$ contact interactions were used,
 another set of azimuthal asymmetries in combination
with forward-backward asymmetries were defined, which are not present in
the present case, and would therefore signal the presence of four-point
interactions.}.

In the next section we discuss numerical evaluation of the cross
sections and asymmetries, and demonstrate  how information using
more than one observable, or one observable, but different polarization
choices can be used to disentangle the different anomalous couplings. We
will also study the numerical limits that can be put on the couplings at
a linear collider.

\section{Numerical Calculations}

We now evaluate various observables and their sensitivities 
for a linear collider operating at
$\sqrt{s}=500$ GeV. We assume that longitudinal beam polarizations of
$P_L=\pm 0.8$ and $\overline P_L=\pm 0.6$ can be reached, and that
rotating the spins to point in the transverse direction will not entail
any loss of polarization. With this choice of individual polarizations,
the factor $1-P_L\overline P_L$, occurring in the expression for the
cross section, 
is 0.52 or 1.48 depending on whether the electron and positron have 
like-sign or unlike-sign polarizations. (We take the sign of
polarization to be positive for right-handed polarization). The quantity
$P_L^{\rm eff}$, defined in eq. (\ref{pleff}), which 
appears in various expressions is then 0.385 or 0.946
in the two cases of like-sign and unlike-sign polarizations.

In case of transverse polarization, we assume $\delta=0$ corresponding
to the simplest configuration of the electron and positron spins.

We have chosen $m_H=120$ GeV for the main part of our calculations. 
We comment later on
the results for larger Higgs masses.

We have made use of the following values of other parameters: $M_Z=91.19$ GeV, 
$\alpha(m_Z) = 1/128$, $\sin^2\theta = 0.22$. 
For studying the sensitivity of the linear collider, we have assumed an 
integrated luminosity of $L \equiv\int \mathcal L dt = 500\, {\mathrm fb}^{-1}$.

\subsection{Cross section}

The simplest observable is the total rate that can be used to determine 
some combination of anomalous couplings. If we  integrate  the 
differential cross section with respect to polar and azimuthal 
angle over the full ranges, we would get a combination 
of the couplings $\Re \Delta a_Z$, $\Re b_Z$, $\Re a_{\gamma}$ and 
$\Re b_{\gamma}$. 
Different combinations of these same couplings enter the 
unpolarized cross section and cross sections 
with same-sign or opposite-sign polarizations of the 
beams. 
Transverse polarization, on the other hand, 
 gives the same combination of the couplings
as in the unpolarized case.

The anomalous part of the cross section in eq. (\ref{partialcs}) 
can be written as
\be
\begin{array}{r}
\sigma_L(\theta_0)-\sigma_L^{\rm SM}(\theta_0)= \sigma_L^{\rm SM}(\theta_0)
\dsp
\left[2\left( \Re \Delta a_Z +
\frac{2\sqrt{s}q^0}{2m_Z^2
+ \left( 1 - \frac{1}{3} \cos^2 \theta_0\right) \vert
\vec q \vert^2}
\Re b_Z \right) \right.\\
\dsp
+ \left.
\frac{(g_V^e - g_A^e P_L^{\rm eff})}
{  
(g_V^{e2} + g_A^{e2} - 2 g_V^{e}g_A^{e} P_L^{\rm
eff} )
} \frac{K_\gamma}{K_{\mathrm SM}}
\left( \Re a_\gamma + 
\frac{2\sqrt{s}q^0}{2m_Z^2
+ \left( 1 - \frac{1}{3} \cos^2 \theta_0\right) \vert
\vec q \vert^2}\Re b_\gamma  \right)
\right]
\end{array}
\ee
It can be seen that for fixed cut-off, measuring  the cross section 
for two different polarization combinations can determine the 
two combinations of two anomalous couplings each.
We define the following two combinations:
\be
 c_Z= 2\left( \Re \Delta a_Z +
\frac{2\sqrt{s}q^0}{2m_Z^2
+ \left( 1 - \frac{1}{3} \cos^2 \theta_0\right) \vert
\vec q \vert^2}
\Re b_Z \right)
\ee
and 
\be
c_{\gamma} = 
\frac{2 g_V^e\sin^22\theta_W}{g_V^{e2} + g_A^{e2}}\frac{s-m_Z^2}{s} 
\left( \Re a_\gamma +
\frac{2\sqrt{s}q^0}{2m_Z^2
+ \left( 1 - \frac{1}{3} \cos^2 \theta_0\right) \vert
\vec q \vert^2}\Re b_\gamma  \right).
\ee
Further, using the same combinations of polarizations, 
$c_Z$ and $c_{\gamma}$ can again be determined for a different value of
cut-off $\theta_0$. This would give two equations for each of $c_Z$ and
$c_{\gamma}$. It would then be possible to determine all four of $ \Re
\Delta a_Z$, $\Re b_Z$, $ \Re a_\gamma $ and $\Re b_\gamma$ independent
of one another. 
\begin{center}
 \begin{figure}[h]
\centering
 \includegraphics[scale=0.9]{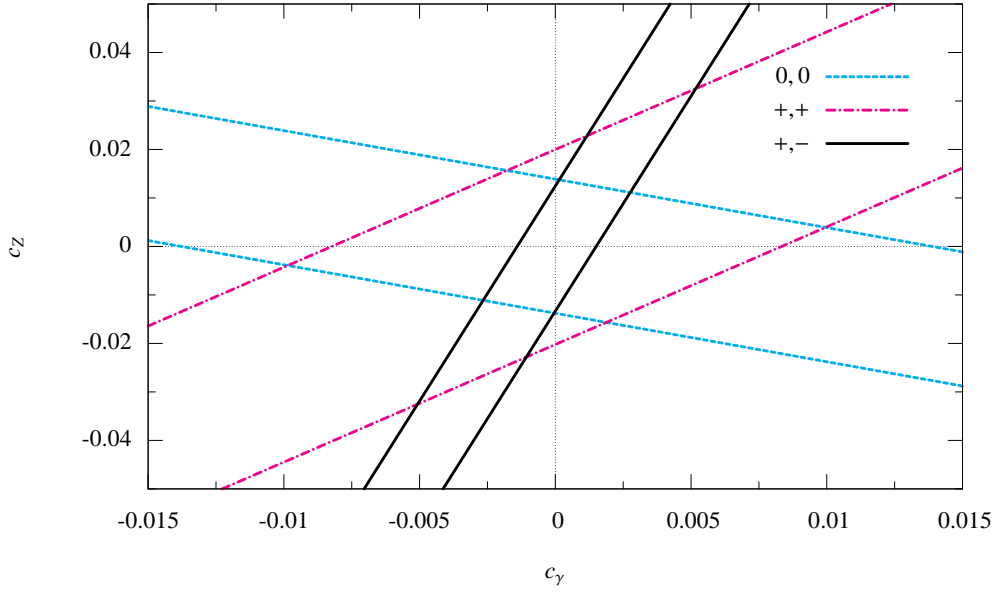}
 \caption{The region in the $c_\gamma - c_Z$ plane accessible at the 
$95$\% CL with cross sections with different beam polarization
configurations for integrated luminosity $L = 
500$ fb$^{-1}$. $0,0$, $+,+$ and $+,-$ stand for the cases of zero, 
like-sign and opposite-sign $e^-$ and $e^+$ polarizations. The cut-off
$\theta_0$ is taken to be $\pi/16$.}
\label{fig:cgiczi}\end{figure}
\end{center}
Fig. \ref{fig:cgiczi} shows the 95\% CL constraints in the $c_\gamma -
c_Z$ plane from polarization combinations $(P_L,\overline P_L)$
of $(0,0)$, $(0.8,+0.6)$ and $(0.8,-0.6)$, using a cut-off $\theta_0 =
\pi/16$. 
The lines correspond to the solutions of the equation 
\be
\vert \sigma_L(\theta_0)-\sigma_L^{\rm SM}(\theta_0) \vert 
= 2.45 \sqrt{\sigma_L^{\rm SM}(\theta_0)/L}
\ee
for the three polarization combinations. 
The best simultaneous limits  on $c_\gamma$ and $c_Z$ are obtained using
a combination of unpolarized beams 
and longitudinally polarized beams with opposite signs, viz.,
\be  
\vert \Re c_\gamma \vert\leq 0.00271, \; \vert\Re c_Z\vert\leq 0.0137.
\ee
The individual limits that can be obtained keeping one coupling to be
nonzero at a time and setting the rest to be zero are shown in Table
\ref{tab:csind}.
\begin{center}
\begin{table}[h]
\centering
\begin{tabular}{|l|c|c|c|c|}
\hline
&$\vert\agr\vert$ & $\vert\azr\vert$ &$ \vert\bgr\vert$&$
\vert\bzr\vert$ \\ \hline
Unpolarized &
0.0705 & 0.00553 & 0.0149  & 0.00117 \\
$P_L=0.8,\,\overline P_L=+0.6$ &
  0.0423 &  0.00805; & 0.00890 & 0.00169 \\
$P_L=0.8,\,\overline P_L=-0.6$ &
0.00741 &  0.00516 & 0.00156  & 0.00109  \\
\hline
\end{tabular}
\caption{Individual 95\% CL limits on the couplings $\agr$, $\azr$,
$\bgr$, $\bzr$ obtained from the cross section for a cut-off
$\theta_0=\pi/16$ for different beam polarization combinations.}
\label{tab:csind}
\end{table}
\end{center}
A direct procedure would of course be to determine all four couplings by
solving four simultaneous equations obtained by using two combinations
of polarization, each for two values of cut-off. Applying this approach for
polarization combinations $P_L=\overline P_L = 0$ and
 $(P_L,\overline P_L ) = (0.8, - 0.6)$,
and the cut-off
values $\theta_0 = \pi/16$ and $\theta_0 = \pi/4$, we find the 95\% CL limits
of 
\be\label{eq:simab}
\begin{array}{llll}
\vert\agr\vert \leq 0.320; & \vert\azr\vert \leq 0.128; & \vert\bgr\vert
\leq 0.0721; & \vert\bzr\vert \leq
0.0287.
\end{array}
\ee
The two other polarization combinations, viz., $P_L=\overline P_L = 0$
or $(P_L,\overline P_L ) = (0.8, + 0.6)$ used with $(P_L,\overline P_L )
= (0.8, - 0.6)$, give worse limits than these.

\subsection{Forward-backward Asymmetry}

As can be seen from eq. (\ref{fbasyexp}), the forward-backward asymmetry 
defined in eq. (\ref{fbasy})
can be a probe of the combination of imaginary part of the 
couplings $\Im \tilde{b}_Z$ and $\Im \tilde{b}_{\gamma}$. 
We examine the accuracy to which this combination can be determined. 
The limits which can be placed at the 95\% CL 
on the two parameters 
contributing to the asymmetry 
is given by equating the asymmetry to $2.45/\sqrt{N_{\mathrm SM}}$, where 
$N_{\mathrm SM}$ is the number of SM events. This leads to the relation
\be\label{fbasylim}
\vert A_{\mathrm FB} \vert = \frac{2.45}{\sqrt{L \sigma^{\mathrm SM}_L}},
\ee
where $L$ is the integrated luminosity.

\begin{center}
 \begin{figure}[h]
\centering
  \includegraphics[scale=0.9]{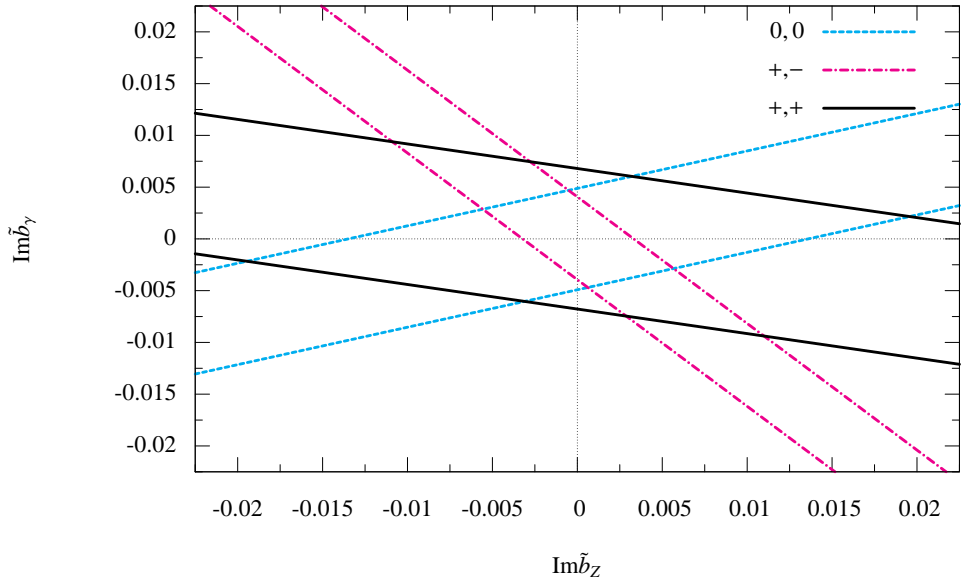}
 \caption{The region in the $\bzti -  \bgti$ plane accessible at the 
$95$\% CL with forward-backward asymmetry with different beam
polarization configurations for integrated luminosity $L = 
500$ fb$^{-1}$. $0,0$, $+,+$ and $+,-$ stand for the cases of zero,
like-sign and opposite-sign $e^-$ and $e^+$ polarizations.}
\label{btzibtgi}\end{figure} 
\end{center}

We show in Fig. \ref{btzibtgi} a plot of the relation eq.
(\ref{fbasylim}) in the space of the couplings involved for unpolarized
beams, and for the two combinations of longitudinal polarizations $(P_L,
\overline P_L)\equiv (0.8,+0.6)$, denoted by $(+,+)$ and $(P_L,
\overline P_L)\equiv (0.8,-0.6)$, denoted by $(+,-)$.
The intersection of the lines corresponding to any two combinations gives 
a closed region which is the allowed region at 95\% CL. 
The best simultaneous limits are 
obtained by considering the region enclosed by the
intersections of the lines corresponding to $P_L=\overline P_L = 0$ and 
$(P_L,\overline P_L) = (0.8, -0.6)$. These limits are
\be 
\vert\bgti\vert \leq 4.69\cdot 10^{-3}; \; \vert\bzti\vert 
\leq 5.61 \cdot 10^{-3}.
\ee 
Individual limits
on the two couplings obtained from the forward-backward asymmetry 
by setting one coupling to zero at a time for
the three polarization combinations are shown in Table \ref{tab:fbind}.

\begin{center}
\begin{table}
\centering
\begin{tabular}{|l|c|c|}
\hline
&$\vert\bgti\vert$ & $\vert\bzti\vert$
\\ \hline
Unpolarized &
0.00392 & 0.0108 \\
$P_L=0.8,\,\overline P_L=+0.6$ &
  0.00543 &  0.0229\\
$P_L=0.8,\,\overline P_L=-0.6$ &
0.00320 &  0.00262 \\
\hline
\end{tabular}
\caption{Individual 95\% CL limits on the couplings $\bgti$, $\bzti$,
obtained from the forward-backward asymmetry for a cut-off
$\theta_0=\pi/16$ for different beam polarization combinations.}
\label{tab:fbind}\end{table}
\end{center}
It can be seen that the limit is improved
considerably in the case of opposite-sign polarizations as compared to
unpolarized beams for $\Im \tilde b_Z$, but only marginally in case of
$\Im \tilde b_\gamma$. Like-sign polarizations make the limits worse.

\subsection{Azimuthal asymmetries} 

Transversely polarized beams can in pri\-nci\-ple pro\-vide more
info\-rma\-tion thr\-ough the 
azimuthal angular distribution which has terms dependent on
$\sin^2\theta$ $ \sin2\phi$ and 
$\sin^2 \theta \cos2\phi$. We can construct observables which isolate
these terms. 

\smallskip
\noindent
\textbf{(a)} The $\sin^2\theta\sin2\phi$ term

This term, which has a coefficient denoted by $E$ in the expression of
eq. (\ref{sigtran}), can be isolated using the asymmetry $A_T(\theta_0)$
defined in eq. (\ref{atfb}), when $\delta = 0$. Since $E_T^{\mathrm SM}$ and
$E_T^{Z}$ are vanishing, this asymmetry uniquely determines 
$E_T^{\gamma}$, and hence the coupling $\Im a_\gamma$. This coupling
cannot be determined without transverse polarization.

We can also choose to evaluate the expectation value of any operators which
are odd functions of $\sin 2\phi$.
We have chosen the three operators 
 sign($\sin2\phi$) whose expectation value corresponds to the asymmetry
$A_T$, $\sin2\phi$ and $\sin^32\phi$.  
The $95$\% CL limit that can 
be placed on $\Im a_\gamma$ was determined for each operator $O$ using
\be\label{limit1}
\vert \Im a_\gamma \vert \leq
1.96\frac{\sqrt{\langle O^2\rangle }}{\langle O \rangle_1\sqrt{L\sigma_T^{SM}}},
\ee
where $\langle O\rangle_1$ is expectation value for unit value of the 
coupling.
Table \ref{tab:agi} shows limits on the $\vert \Im a_{\gamma}\vert$ at the 
$95\%$ confidence level for various Higgs masses. 
\begin{table}[h]
\centering
\begin{tabular}{|c|c|c|c|}
\hline
Operators & $M_H=120$ GeV&  $M_H=200$ GeV & $M_H=300$ GeV \\
\hline
sign($\sin{2\phi}$) &0.0409 & 0.0522 & 0.101\\
\hline
$\sin{2\phi}$       &0.0368 & 0.0470 & 0.0913\\
\hline
$\sin^3{2\phi}$     &0.0388 & 0.0495 & 0.0963\\
\hline
\end{tabular} 
 \caption{Limits of $\Im a_{\gamma}$ for the various Higgs masses}
\label{tab:agi}\end{table}
It is seen that the best limits are obtained using the operator $\sin
2\phi$.

\smallskip
\noindent
\textbf{(b)} The $\sin^2\theta\cos2\phi$ term

The coefficient of the $\cos2\phi$ term, viz., $D_T$, is associated with 
$\azr$ and $\Re a_{\gamma}$. So, by suitably defining an asymmetry 
corresponding to $\cos2\phi$, we can probe a linear combination of 
$\Re \Delta a_Z$  and $\Re a_{\gamma}$. The asymmetry $A^{'T}$ defined in eq.
(\ref{atfb2}) serves the purpose.
This asymmetry does not 
vanish for the SM. Hence we use the following expression for determining 
the limit on the linear combination of couplings at
$95$\% CL.
\be\label{limit2}
 \vert A^{'T} - A^{'T\,\mathrm SM}\vert\leq 2.45 
	\frac{\sqrt{1-(A^{'T\,{\mathrm SM}})^2}}{\sqrt{L
	\sigma^{\mathrm SM}}},
\ee
where $A^{\mathrm SM}$ is the value of the asymmetry in SM.
\comment{
We found this asymmetry to be 
\be\label{A}
 A=-1.142\Re a_{\gamma}-0.399\Re \Delta a_Z
\ee

{\large \bf This has to be set right}
The constraint on this combination of couplings, with the $95$\% 
confidence level (CL) with Higgs mass taken to be $120$ GeV and polar 
angle $\theta$ integrated all over range, is found to be: 
\be\label{limitA}
 -1.142\Re a_{\gamma}-0.399\Re \Delta a_Z\leq 1.67\times 10^{-2}
\ee
}
Since this asymmetry is proportional to the product $P_T\overline
P_T$, changing the sign of polarization will only give a change of sign
of the asymmetry. It is thus not possible to obtain two different
combinations of the couplings $\Re a_{\gamma}$ and $\Re \Delta a_Z$ as in the
earlier case of longitudinal polarization. However, it would be possible
to obtain simultaneous limits on these couplings by choosing two
different cut-offs
on the azimuthal angle $\phi$, which would give two equations. We have
not attempted this in the present work.

The individual limits using $A^{'T}$
on $\Re a_{\gamma}$ and $\Re \Delta a_Z$, each taken
nonzero by turns,  
are
\be 
\vert\agr\vert \leq 0.334, \; \vert\azr \vert\leq 0.0270
\ee
It is seen that the limit on $\agr$ is not an improvement over the one
shown in eq. (\ref{eq:simab}),
obtained using the partial cross sections
with longitudinal polarization. The limit on $\azr$
is a considerable improvement over the limit in eq. (\ref{eq:simab}),
though it is much worse than the individual limit obtained using
even unpolarized beams (Table \ref{tab:csind}).

\section{Conclusions and discussion}

We have obtained angular distributions for the process $e^+e^- \to ZH$
in the presence of anomalous $\gamma ZH$ and $ZZH$ couplings to linear
order in these couplings in the presence of longitudinal and transverse
beam polarizations. We have then looked at observables and asymmetries
which can be used in combinations to disentangle the various couplings
to the extent possible. 
We have also obtained the sensitivities of these observables and
asymmetries to the various couplings for a definite configuration of the
linear collider.

In certain cases where the contribution of a
coupling is suppressed due to the fact that the vector coupling of the
$Z$ to $e^+e^-$ is numerically small, longitudinal polarization helps to
enhance the contribution of this coupling. As a result, longitudinal
polarization improves the sensitivity.
The main advantage of transverse polarization is that it helps to
determine $\agi$ independent of all other couplings through the
$\sin^2\theta\sin 2\phi$ term. 
It is not possible to constrain $\agi$ without transverse polarization. 
Another advantage that transverse polarization offers, though not
as compelling, is the
determination and a combination of
the couplings $\agr$ and $\azr$ independently of the couplings $\bgr$
and $\bzr$ through the $\sin^2\theta\cos 2\phi$ term. 
It is of course
possible to measure the couplings $\agr$ and $\azr$ independently of the
couplings $\bgr$       
and $\bzr$ using unpolarized or longitudinally polarized beams. However,
transverse polarization enables this to be done  using a convenient
azimuthal asymmetry. In the case of $\azr$, this procedure proved to be
more sensitive than the one determining simultaneous limits 
employing cross sections for 
two combinations of longitudinal polarization and two different cut-offs
in the polar angle. 

We find that with a linear collider operating at a c.m. energy of 500
GeV with the capability of 80\% electron polarization and 60\% positron
polarization with an integrated luminosity of 500 fb$^{-1}$, using the
simple cross section and asymmetry measurements described above it would be
possible to place 95\% CL individual limits 
of the order of few times $10^{-3}$ or better on all couplings taken 
nonzero one at a time with use of an appropriate combination ($P_L$ and
$\overline P_L$ of opposite signs) of longitudinal beam polarizations.
Polarization gives an improvement in sensitivity 
by a factor of 5 to 10 as compared to the
unpolarized case for the real parts of $\gamma ZH$ couplings, and
the imaginary parts of $ZZH$ couplings. 
The use of polarization also enables simultaneous determination (without
any coupling being assumed zero) of all couplings 
which appear in the differential cross section, viz., $\azr$, $\agr$,
$\agi$, $\bzr$, $\bgr$, $\bgti$ and $\bzti$.
 The simultaneous limits are, as expected, less stringent, of
the order of $0.1-0.3$ for $\agr$ and $\azr$, and of the order of
$0.03-0.07$ on
$\bgr$ and $\bzr$.  The simultaneous limits on the CP-violating
couplings $\bgti$ and $\bzti$ are a little better, being respectively
$5\cdot 10^{-3}$ and $6\cdot 10^{-3}$. Transverse polarization enables
the determination of $\agi$ independent of all other couplings, with a
possible 95\% CL limit of the order of $10^{-2}$. 
With transverse polarization a
combination of $\agr$ and $\azr$ can be determined independent of
all other couplings with the help of an azimuthal asymmetry. 
From this combination, individual limits possible
on them are respectively $0.33$ and $2.7\cdot 10^{-2}$.
While the former is comparable to the simultaneous limit obtained using
longitudinal polarization, the latter is an improvement by an order of
magnitude.

It is appropriate to compare our results with those in works using the
same paramterization as ours for the anomalous coupling and with an
approach similar to ours. Ref. \cite{han} deals with CP-violating $ZZH$
couplings,
and it is possible to compare the 95\% CL limits obtained in Sec. 5.2 using the
forward-backward asymmetry of the $Z$ with the corresponding limits in
\cite{han}. With identical values of $\sqrt{s}$ and integrated
luminosity, ref. \cite{han} quotes limits of 0.019 and 0.0028 for
$\bzti$, respectively for unpolarized and longitudinally polarized beams
with opposite-sign $e^+$ and $e^-$ polarizations. The corresponding
numbers we have are 0.011 and 0.0026. The agreement is thus good,
considering that ref. \cite{han} employs additional experimental cuts,
which could reduce the nominal sensitivity.
The papers in \cite{biswal} also deal only with anomalous $ZZH$
couplings, and quote $3\sigma$ limits on the couplings. However, the
second paper in \cite{biswal} considers different luminosity options for
different polarization combinations. The $3\sigma$ limit they quote for
$\bzti$ is  $0.064$ for unpolarized beams, and $0.0089$ for
polarized beams. After correcting for the CL limit of $1.96\sigma$
which we use, their limits are still worse by a factor of 2 to 4. This
could be attributed to the stringent kinematic cuts imposed by them, and
to the different luminosity choice in the case of polarized beams.
Similarly, the limits quoted in \cite{biswal} for $\azr$ and $\bzr$ 
are worse by a
factor of about 2 or 3 in the unpolarized as well as polarized cases.
As for the case of $\gamma ZH$ couplings, comparison with earlier work
is not easy because of the different approach to paramterization of
couplings. Also, there is no work dealing in transverse polarization
with which we could make a comparison.

In the above, we have assumed a Higgs mass of 120 GeV. For larger values
of $m_H$, for larger Higgs masses, we find decreased sensitivities. 
We have not studied the sensitivities at higher centre-of-mass energies.
However, it is conceivable that the contribution of the anomalous couplings
$b_{\gamma,Z}$ and $\tilde b_{\gamma,Z}$ which come with 
momentum-dependent tensors
will increase with energy and improve the sensitivity.

Though we have used SM couplings for the leading contribution of Fig.
\ref{fig:vvhptgraph}, as mentioned earlier, the analysis needs only
trivial
modification when applied to a model like MSSM or a multi-Higgs-doublet
model, and will be useful in such extensions of SM.

We have not included the decay of the $Z$ and the Higgs boson in our
analysis. For now, one could simply divide our limits by the square
root of the branching ratios and detection efficiencies. 
Including these decays will entail some loss of efficiency. On
the other hand, making use of the decay products of the $Z$ would also
give access to more variables, which might help one to disentangle
further the couplings which could not be easily disentangled in our
analysis. In particular, $\azi$, $\bgi$, $\bzi$, 
$\bgtr$ and $\bztr$, which do not appear in the
$Z$ angular distributions, will most likely be accessible in the
distributions of $Z$ decay products.
Work on this is under progress.

We have not considered scenarios with an extra neutral gauge boson $Z'$.
While it is straightforward to include a $Z'$ in our analysis, the number
of couplings would be much larger and difficult to disentangle without
studying the $s$ dependence of the cross section or asymmetries.

One should also investigate  the effect experimental cuts would
have on the accuracy of the determination of the couplings. 
One should keep in mind the possibility that radiative corrections can
lead to quantitative changes in the above results (see, for example, 
\cite{comelli}).
While these
practical questions are not addressed in this work, 
we feel that the interesting new features we found would make it
worthwhile to address them in future.

\noindent{\bf Acknowldgement} We thank Sudhansu Biswal for suggestions
and a careful
reading of the manuscript.

\comment{
\bibitem{kile}
  J.~Kile and M.~J.~Ramsey-Musolf,
  arXiv:0705.0554 [hep-ph].
} 

\bibitem{basdr}
B.~Ananthanarayan and S.~D.~Rindani,
Phys.\ Lett.\ B {\bf 606}, 107 (2005) 
[arXiv:hep-ph/0410084];
JHEP {\bf 0510}, 077 (2005)
[arXiv:hep-ph/0507037];
 
 \bibitem{lepto}
S.~D.~Rindani,
Phys.\ Lett.\ B {\bf 602}, 97 (2004)
[arXiv:hep-ph/0408083].
\comment{
\bibitem{chen}
  C.~M.~J.~Chen, J.~W.~J.~Chen and W.~Y.~P.~Hwang,
  Phys.\ Rev.\  D {\bf 50}, 4485 (1994).

\bibitem{mahlon}
  G.~Mahlon and S.~J.~Parke,
  Phys.\ Rev.\  D {\bf 74}, 073001 (2006)
  [arXiv:hep-ph/0606052].
}  
\bibitem{rizzo}
T.~G.~Rizzo,
JHEP {\bf 0302}, 008 (2003)
[arXiv:hep-ph/0211374];
  JHEP {\bf 0308}, 051 (2003)
  [arXiv:hep-ph/0306283];
%
J.~Fleischer, K.~Kolodziej and F.~Jegerlehner,
Phys.\ Rev.\ D {\bf 49}, 2174 (1994);
%
M.~Diehl, O.~Nachtmann and F.~Nagel,
Eur.\ Phys.\ J.\ C {\bf 32}, 17 (2003)
[arXiv:hep-ph/0306247];
%
S.~Y.~Choi, J.~Kalinowski, G.~Moortgat-Pick and P.~M.~Zerwas,
Eur.\ Phys.\ J.\ C {\bf 22}, 563 (2001)
[Addendum-ibid.\ C {\bf 23}, 769 (2002)]
[arXiv:hep-ph/0108117];
A.~Bartl, K.~Hohenwarter-Sodek, T.~Kernreiter and H.~Rud,
Eur.\ Phys.\ J.\ C {\bf 36}, 515 (2004)
[arXiv:hep-ph/0403265];
%
J.~Kalinowski,
arXiv:hep-ph/0410137;
%
P.~Osland and N.~Paver,
arXiv:hep-ph/0507185;
%
A.~Bartl, H.~Fraas, S.~Hesselbach, K.~Hohenwarter-Sodek, T.~Kernreiter
and G.~Moortgat-Pick,
JHEP {\bf 0601}, 170 (2006)
[arXiv:hep-ph/0510029];
%
S.~Y.~Choi, M.~Drees and J.~Song,
 JHEP {\bf 0609}, 064 (2006)
  [arXiv:hep-ph/0602131];
%
  K.~Huitu and S.~K.~Rai,
  Phys.\ Rev.\  D {\bf 77}, 035015 (2008)
  [arXiv:0711.4754 [hep-ph]].

\bibitem{basdrDR}
  B.~Ananthanarayan and S.~D.~Rindani,
  Eur.\ Phys.\ J.\  C {\bf 46}, 705 (2006);
  [arXiv:hep-ph/0601199].
%
  Eur.\ Phys.\ J.\  C {\bf 56}, 171 (2008)
  [arXiv:0805.2279 [hep-ph]].

\bibitem{poulose} 
  P.~Poulose and S.~D.~Rindani,
  Phys.\ Lett.\  B {\bf 383}, 212 (1996)
  [arXiv:hep-ph/9606356];
  Phys.\ Rev.\  D {\bf 54}, 4326 (1996)
  [Erratum-ibid.\  D {\bf 61}, 119901 (2000)]
  [arXiv:hep-ph/9509299];
  Phys.\ Lett.\  B {\bf 349}, 379 (1995)
  [arXiv:hep-ph/9410357];
 F.~Cuypers and S.~D.~Rindani,
  Phys.\ Lett.\  B {\bf 343}, 333 (1995)
  [arXiv:hep-ph/9409243].
  D.~Choudhury and S.~D.~Rindani,
  Phys.\ Lett.\  B {\bf 335}, 198 (1994)
  [arXiv:hep-ph/9405242];
  S.~D.~Rindani,
  Pramana {\bf 61}, 33 (2003)
  [arXiv:hep-ph/0304046].

\bibitem{form} 
  J.~A.~M.~Vermaseren,
  arXiv:math-ph/0010025.

\bibitem{hikasa} K.-i.~Hikasa,
Phys.\ Rev.\ D {\bf 33}, 3203 (1986).
 
\bibitem{sdrCP} See, for example, S.D.~Rindani, 
  Pramana {\bf 49}, 81 (1997);
  Pramana {\bf 45}, S263 (1995)
  [arXiv:hep-ph/9411398];
  G.~C.~Branco, L.~Lavoura and J.~P.~Silva,
{\it  Oxford, UK: Clarendon (1999) 511 p}

\bibitem{comelli}
P.~Ciafaloni, D.~Comelli and A.~Vergine,
JHEP {\bf 0407}, 039 (2004).
[arXiv:hep-ph/0311260].

\end{document}